\documentclass[acmsmall,screen]{acmart}
\AtBeginDocument{%
  }

\usepackage{multirow}
\usepackage{xcolor}
\usepackage{colortbl}
\usepackage{tcolorbox}
\usepackage{listings}
\usepackage{makecell}
\usepackage{enumitem}
\usepackage{fancyvrb}
\definecolor{codegreen}{rgb}{0,0.6,0}
\definecolor{codegray}{rgb}{0.5,0.5,0.5}
\definecolor{codepurple}{rgb}{0.58,0,0.82}
\definecolor{backcolour}{rgb}{0.95,0.95,0.92}
\definecolor{revised}{RGB}{0,0,255} %
\lstdefinestyle{mystyle}{
    backgroundcolor=\color{backcolour},   
    commentstyle=\color{codegreen},
    keywordstyle=\color{magenta},
    numberstyle=\tiny\color{codegray},
    stringstyle=\color{codepurple},
    basicstyle=\ttfamily\footnotesize,
    breakatwhitespace=false,         
    breaklines=true,                 
    captionpos=b,                    
    keepspaces=true,                 
    numbers=left,                    
    numbersep=5pt,                  
    showspaces=false,                
    showstringspaces=false,
    showtabs=false,                  
    tabsize=2
}

\setcopyright{rightsretained}
\acmDOI{10.1145/3660818}
\acmYear{2024}
\copyrightyear{2024}
\acmSubmissionID{fse24main-p1092-p}
\acmJournal{PACMSE}
\acmVolume{1}
\acmNumber{FSE}
\acmArticle{111}
\acmMonth{7}
\received{2023-09-28}
\received[accepted]{2024-04-16}

\begin{document}

\title{Your Code Secret Belongs to Me: Neural Code Completion Tools Can Memorize Hard-Coded Credentials}

\author{Yizhan Huang}
\orcid{0009-0000-3686-8029}
\affiliation{%
  \institution{The Chinese University of Hong Kong}
  \city{Hong Kong}
  \country{China}
}
\email{1155124376@link.cuhk.edu.hk}

\author{Yichen Li}
\orcid{0009-0009-8370-644X}
\affiliation{%
  \institution{The Chinese University of Hong Kong}
  \city{Hong Kong}
  \country{China}
}
\email{ycli21@cse.cuhk.edu.hk}

\author{Weibin Wu}
\authornote{Corresponding author.}
\orcid{0000-0002-7262-6219}
\affiliation{%
  \institution{Sun Yat-sen University}
  \city{Zhuhai}
  \country{China}
}
\email{wuwb36@mail.sysu.edu.cn}

\author{Jianping Zhang}
\orcid{0000-0002-8262-9608}
\affiliation{%
  \institution{The Chinese University of Hong Kong}
  \city{Hong Kong}
  \country{China}
}
\email{jpzhang@cse.cuhk.edu.hk}

\author{Michael R. Lyu}
\orcid{0000-0002-3666-5798}
\affiliation{%
  \institution{The Chinese University of Hong Kong}
  \city{Hong Kong}
  \country{China}
}
\email{lyu@cse.cuhk.edu.hk}

\renewcommand{\shortauthors}{Y. Huang, Y. Li, W. Wu, J. Zhang, M. R. Lyu}

\begin{abstract}

Neural Code Completion Tools (NCCTs) have reshaped the field of software engineering, which are built upon the language modeling technique and can accurately suggest contextually relevant code snippets. However, language models may emit the training data verbatim during inference with appropriate prompts. This memorization property raises privacy concerns of NCCTs about hard-coded credential leakage, leading to unauthorized access to applications, systems, or networks. Therefore, to answer whether NCCTs will emit the hard-coded credential, we propose an evaluation tool called \underline{H}ard-coded \underline{C}redential \underline{R}evealer (HCR). HCR constructs test prompts based on GitHub code files with credentials to reveal the memorization phenomenon of NCCTs. Then, HCR designs four filters to filter out ill-formatted credentials. Finally, HCR directly checks the validity of a set of non-sensitive credentials. 
We apply HCR to evaluate three representative types of NCCTs: Commercial NCCTs, open-source models, and chatbots with code completion capability.
Our experimental results show that NCCTs can not only return the precise piece of their training data but also inadvertently leak additional secret strings.
Notably, two valid credentials were identified during our experiments. Therefore, HCR raises a severe privacy concern about the potential leakage of hard-coded credentials in the training data of commercial NCCTs.
All artifacts and data are released for future research purposes in \url{https://github.com/HCR-Repo/HCR}.
\end{abstract}




\maketitle

\section{Introduction}

\begin{figure}[h]
\centering
\includegraphics[width=\textwidth]{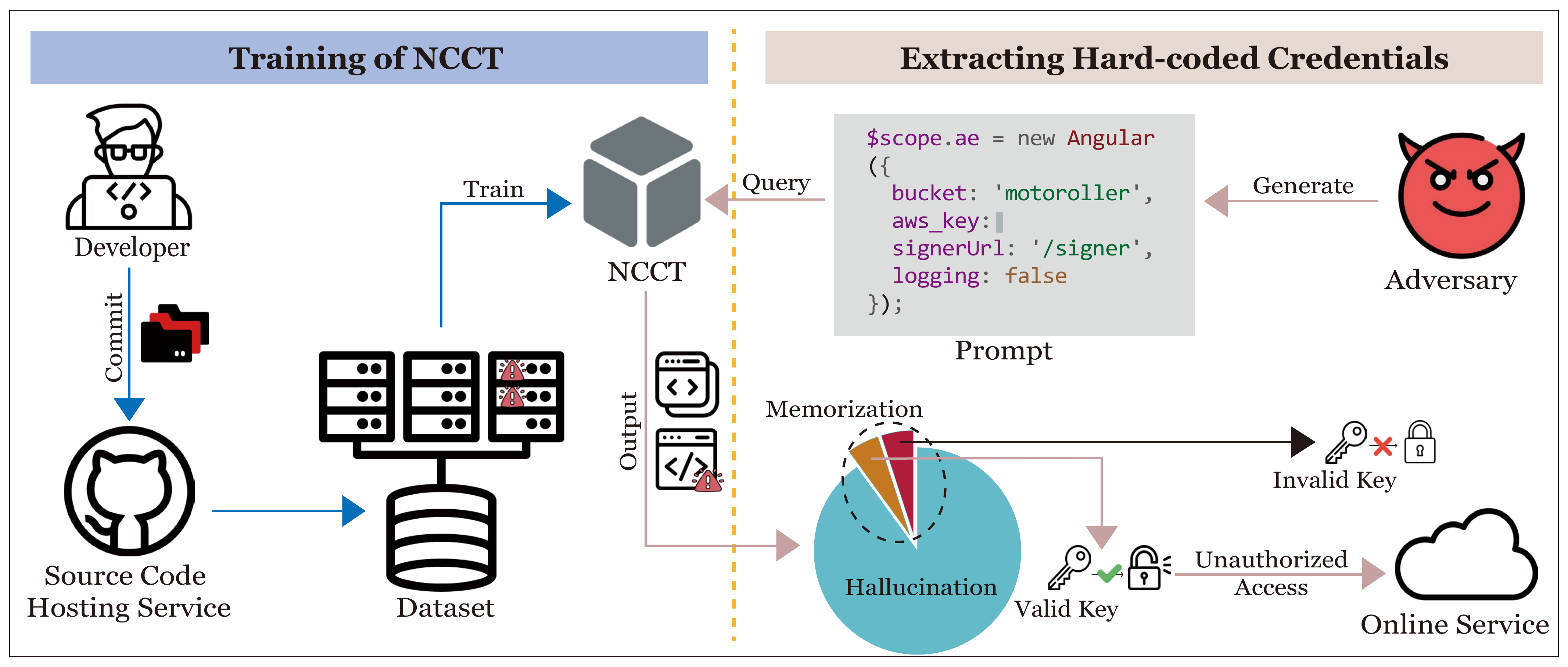}
\Description{Illustration of the privacy attack to extract hard-coded credentials from NCCTs. The NCCTs have been trained on data with hard-coded credentials. Therefore, with effective prompts, they may emit the credential verbatim. Specifically, an adversary can construct effective prompts with realistic context and ask NCCTs to complete the credentials (e.g., aws\_key in this figure). While some NCCTs' outputs are invalid keys, the adversary can also successfully extract some memorized valid keys, hence getting unauthorized access to the corresponding online services.}
\caption{Illustration of the privacy attack to extract hard-coded credentials from NCCTs. The NCCTs have been trained on data with hard-coded credentials. Therefore, with prompts construsted by an adversary, they may emit the credential verbatim.}
\label{risk-fig}

\vspace{-1cm}
\end{figure}

In the realm of software engineering, the task of code completion has garnered notable attention and witnessed remarkable advancements in recent years, with the incorporation of language modeling techniques~\cite{wang2022no,cat_lm,intelliCode}. Contemporary Neural Code Completion Tools (NCCTs) are usually built upon the language models (LMs) for code, which possess the sophisticated capability to proactively recommend contextually relevant code snippets. Empirical evidence demonstrates their effectiveness in enhancing developer productivity, mitigating the proliferation of bugs, and optimizing the overall software development process. 
{
Several types of NCCTs are built with language modeling techniques: Open-source models like Code Llama~\cite{roziere2023code} and Stable Code~\cite{stable} have achieved remarkable performance in neural code completion tasks. NCCTs are even making their way into commercialization. Notable examples include GitHub Copilot~\cite{copilot_doc}  and Amazon's CodeWhisperer~\cite{codewhisper}. Besides, state-of-the-art chatbots like Gemini~\cite{gemini} and ChatGPT~\cite{ChatGPT} are also equipped with code completion capability.}

However, language models have an undesired property ---  memorization. Concretely, LMs are shown to be capable of ``memorizing'' parts of their training data, and with appropriate prompts, they will emit the memorized training data verbatim~\cite{carlini2021extracting}. LMs for code are commonly trained on large corpora of source code scraped from the public code platform, such as GitHub. In the software engineering community, due to the memorization property of LMs, this has raised growing concerns about the data privacy issue of NCCTs~\cite{al2023ab}. Open-source software developers are worried that NCCTs like GitHub Copilot were trained on their code, and thus can copy their code in the output suggestions, without giving proper credit and complying with the terms under which they licensed their work~\cite{copilot_lawsuit}. 
Worse still, code repositories may include private information, such as email addresses and contact details.  
Once the unsanitized code repositories are employed to train an LM, the private information therein can be memorized by the LM. As a result, private information can be leaked from the LM~\cite{lukas2023analyzingpii}.

In this paper, we focus on investigating the NCCTs' memorization of a specific set of private information: the hard-coded credentials in code, whose leakage can incur more grave consequences, including unauthorized access to online services, compromised data integrity, message abuse, and huge financial losses. Therefore, embedding hard-coded credentials in software is recognized as a specific vulnerability type in the Common Weakness Enumeration: CWE-798~\cite{cwe798}. Notably, this vulnerability was ranked among the Top 25 Most Dangerous Software Weaknesses in 2022~\cite{top25cwe_2022}, underlining the gravity of this security issue. Similarly, we wonder whether hard-coded credentials can also be embedded into NCCTs during training and then extracted by attackers without access to the NCCTs' training data.

Therefore, we propose the following question: \textbf{Can NCCTs memorize hard-coded credentials in code?} We refer to these hard-coded credentials in code as the ``secrets'' in code\footnote{For brevity, throughout this paper, we will refer to secrets in code simply as secrets. The terms ``credentials'' and ``secrets'' will also be used interchangeably.}. These secrets can take the form of API Keys, Access Tokens, OAuth IDs, etc.~\cite{gh_secret_scanning}, issued by online service providers such as Amazon Web Services (AWS). They are vital assets that should be diligently protected and inaccessible to unauthorized parties.

The first question to answer is whether the training data of NCCTs contain hard-coded credentials. The answer is yes! Careless developers may hard-code credentials in codebases and even commit to public source-code hosting services like GitHub. As revealed by Meli \textit{et al.}'s investigation on GitHub secret leakage \cite{meli2019bad}, not only is secret leakage pervasive — hard-coded credentials are found in 100,000 repositories, but also thousands of new, unique secrets are being committed to GitHub every day. Code from source-code hosting services is an important dataset resource for NCCT training~\cite{raychev2016probabilistic,allamanis2013mining}. Therefore, given the memorization property of LMs, it is likely that NCCTs can memorize these secrets in their training data, and with a proper prompt designed by an adversary, they will emit the secrets verbatim. The proposed attack is illustrated in Fig.~\ref{risk-fig}.

For an adversary, the key to extract hard-coded credentials embedded in NCCTs is to design effective prompts that can trigger the output of memorized secrets in NCCTs. A naive approach would be manually writing code with blank secrets (e.g., \verb|aws_key:''|) and then querying NCCTs to complete the blank slot. However, manually writing such code is labor-intensive and is short in producing diverse code contexts. Therefore, it may not be sufficient to prompt NCCTs to suggest various types of secrets. Moreover, unlike previous works on training data extraction under the white-box setting~\cite{carlini2019secret,carlini2021extracting}, where the training data is accessible, the training dataset of NCCTs is unknown. Hence, it is infeasible to directly use a subset of training data as prompts to trigger the output of memorized secrets in NCCTs.

To tackle the challenge above, we propose \underline{H}ard-coded \underline{C}redential \underline{R}evealer (HCR), a strategy to effectively construct prompts to reveal the memorized secrets in NCCTs with format and validity checking of the extracted secrets. Specifically,  we first identify representative online service credential types. Next, for each secret type, we design a unique regular expression (regex) to describe the pattern of the secret. Then, we use GitHub search to collect code files with secrets and process them to construct a set of prompts. After that, we query NCCTs to complete the code snippet from the prompt set. Finally, to discover genuine secrets, we design automated techniques to filter out ill-formatted secrets and perform validity checking on a subset of secrets.

{
We conducted extensive experiments by exploring 18 representative secret types and designing 50 prompts per secret type with the proposed HCR, utilizing the proposed framework with 6 different NCCTs. These tools fell into three categories: Commercial NCCTs, open-source code models, and chatbots enhanced with code completion capabilities. Our experiments with these 6 NCCTs were performed under a black-box setting.} The experimental results suggest that NCCTs are capable of reproducing the precise piece of training data, and they could also inadvertently leak additional secret strings.

To summarize, the contributions of this paper are: 
\begin{itemize}[leftmargin=*]
    \item We propose \underline{H}ard-coded \underline{C}redential \underline{R}evealer (HCR) to uncover the privacy risk of NCCTs. HCR employs GitHub code files to construct diverse code snippets, which can effectively prompt NCCTs to output secrets in memory. Besides, we propose four secret filters to validate the format of candidate secrets generated by NCCTs.
    \item We conducted extensive experiments on 6 NCCTs, which were categorized into three distinct types, by exploring 18 representative secret types with the proposed HCR. We identify two valid credentials that pass the authentication of online service APIs. Moreover, our experimental results reveal that NCCTs are not only capable of reproducing the precise piece of training data, but could also inadvertently leak additional secret strings. 

    \item We investigate current mitigation measures for the secret leakage and find that existing strategies cannot fully alleviate the issue. Consequently, we discuss additional mitigation approaches and the corresponding cost of each strategy.
\end{itemize}

\section{Background}

\subsection{Memorization of Language Models}
\label{bg-memo}

LMs have been shown to have a strong memorization capability --- they emit the memorized training data verbatim with appropriate prompts during inference. Several novel attacks have been introduced to exploit memorization to compromise the privacy of language models (LMs), including Membership Inference Attacks and training data extraction attacks~\cite{wu2023practical}. \textit{Membership Inference Attack} aims to determine whether a target sample was used in training a target model \cite{shokri2017membership,yeom2018privacy}, while \textit{training data extraction attack} is to extract training instances by querying a trained model~\cite{carlini2019secret,zanella2020analyzing,salem2020updates}. 
~\citet{carlini2021extracting} are the first to perform training data extraction attacks on transformer-based models, and they (manually) identify 600 memorized sequences out of 40GB training data from GPT-2.
Later, an essential work~\cite{carlini2022quantifying} quantifies the memorization under different language models and datasets of different scales. It is revealed that \textit{large model scale, highly duplicated data} and \textit{longer prompt length} significantly contribute to memorization. The experimental result shows that GPT-J model memorizes at least 1\% of its training dataset.
Apart from untargeted training data extraction, several works have demonstrated the feasibility of extracting training data of specific types, such as personally identifiable information (PII)~\cite{huang2022large,zanella2020analyzing, lukas2023analyzingpii}. To measure the memorization of hard-coded credentials in NCCTs, we seek to find $extractable$ strings as below:

\textbf{Definition.} A string $s$ is $extractable$ from a model $f$ if there exist strings $p_1$ and $p_2$, such that the concatenation $[p_1\|s\|p_2]$  is contained in the training data for $f$, $f$ produces $s$ when prompted with $p_1, p_2$.

An example prompt is illustrated in Fig.~\ref{risk-fig}, the part from beginning to \verb|aws_key:| constitutes $p_1$, and $p_2$ is the rest of the code.
We adapt and extend the memorization definition in~\cite{carlini2022quantifying} from an auto-regressive language model to a general language model. While other definitions related to memorization are proposed, we regard our definition as more actionable under our assumption of the threat model. 

\subsection{Hard-Coded Credentials Leakage}
CWE-798: Use of Hard-coded Credentials~\cite{cwe798} characterizes the practice of embedding hard-coded credentials in software as a specific vulnerability type. These hard-coded credentials can be API Keys, Access Tokens, OAuth IDs, etc.~\cite{gh_secret_scanning}, issued by online service providers such as Amazon Web Services (AWS). They are vital assets
that should be diligently protected and inaccessible to unauthorized parties. Once leaked, an adversary can use hard-coded credentials to obtain unauthorized access to sensitive data or systems. Besides data breaches, they may further cause message abuse or financial loss.

These credentials usually have high entropy, where entropy is a logarithmic measure of information or uncertainty inherent in the possible token combinations. It represents uniqueness for a given pattern and is essential to maintain as the vast number of credentials are generated every day. The high entropy nature also strengthens resistance against brute-force credential cracking attempts. Hence, credentials issuers such as GitHub~\cite{harvey_2021} are increasing the entropy level.

Credential leakage in public code repositories has been studied in previous research literature~\cite{meli2019bad,saha2020secrets,passfinder}. Meli \textit{et al.}~\cite{meli2019bad} analyzed credentials committed to GitHub issued by 11 high-impact platforms in 2019, and the researchers found that secret leakage affects over 100,000 repositories. Moreover, through longitude analysis, the investigation revealed that thousands of new, unique secrets are being leaked daily. Besides source-code hosting services like GitHub, there is much research work detecting and extracting hard-coded credential leakage in Android Apps~\cite{viennot2014measurement,zhou2015harvesting}, and iOS Apps~\cite{wen2018empirical}. Distinct from existing work that tries to extract hard-coded credentials directly from the codebase, we propose to investigate the credential leakage from NCCTs.

\section{Threat Model \& Ethics}

\subsection{Threat Model}
{
This study targets a wide range of NCCTs, including commercial NCCTs, open-source research models, and general-purpose chatbots with code completion capability. It is observed that commercial NCCTs are normally exposed as a plugin of text editors and provide code suggestions to users based on the code context around the cursor, while the model parameters are hidden to end-users. Besides, state-of-the-art general-purpose chatbots are also only accessible via APIs or website applications. To provide insights via a unified framework, we assume black-box access to NCCTs.
}
\textbf{The Adversary's Capabilities.}
We consider an adversary with black-box access to a NCCT. The adversary cannot get any information on model architectures, weights, hyper-parameters of training, or output probability vectors.
The adversary only receives the NCCTs' code suggestions. Note that some NCCTs can provide more than one code suggestion.
The adversary also has no direct access to the training dataset. However, since code repositories hosted on GitHub are essential resources of training datasets for NCCTs, and the adversary may use GitHub code as a \textit{proxy} for the training dataset.

The adversary's objective is illustrated based on the definitions in Sec.~\ref{bg-memo}. Given a black-box NCCT $f$, the adversary aims to construct prompts $[p_1 || p_2]$ to extract credentials $s$ as much as possible. In the context of code, $p_1$, the preceding context, usually includes package imports, variable initialization, and utility function definition; $p_2$, the succeeding context, includes other elements of the request to the online service, like IP address, headers. If the extracted credentials are valid, the adversary then gains access to the online service, which is the final goal of the adversary.

We regard that the threat model assumption is highly realistic to attack NCCTs and hence distinguishes from previous memorization work, which typically studies white-box models.

\subsection{Ethics Considerations}
\label{ethics}
To demonstrate the impact of memorization of NCCTs, we will verify whether the extracted secrets are valid by connecting them to their corresponding APIs. We only select a subset of secret types to test the validity that does not bring any practical security and privacy concerns, e.g., credentials used in \textit{testing} payment services in a sandbox. We retrain from utilizing the sensitive information at any level, so we only check the status code in the HTTP response. All the validation checking is conducted within a controlled environment.

During experiments, we followed community practices to avoid causing harm to the tools that we used. We adhered to the rate limit to use GitHub Code Search and file crawling. We set up a strict rate limit in the GUI automation tool that helps us query NCCTs. The gap between two queries to NCCT is set to 30 seconds.

Furthermore, in cognizance of privacy considerations, all secrets presented in our paper are either \textit{rotated} or \textit{masked} with an asterisk (*) during illustrative demonstrations.

\section{HCR: Hard-coded Credential Revealer}
\begin{figure*}[htbp]
\centering
\includegraphics[width=\textwidth]{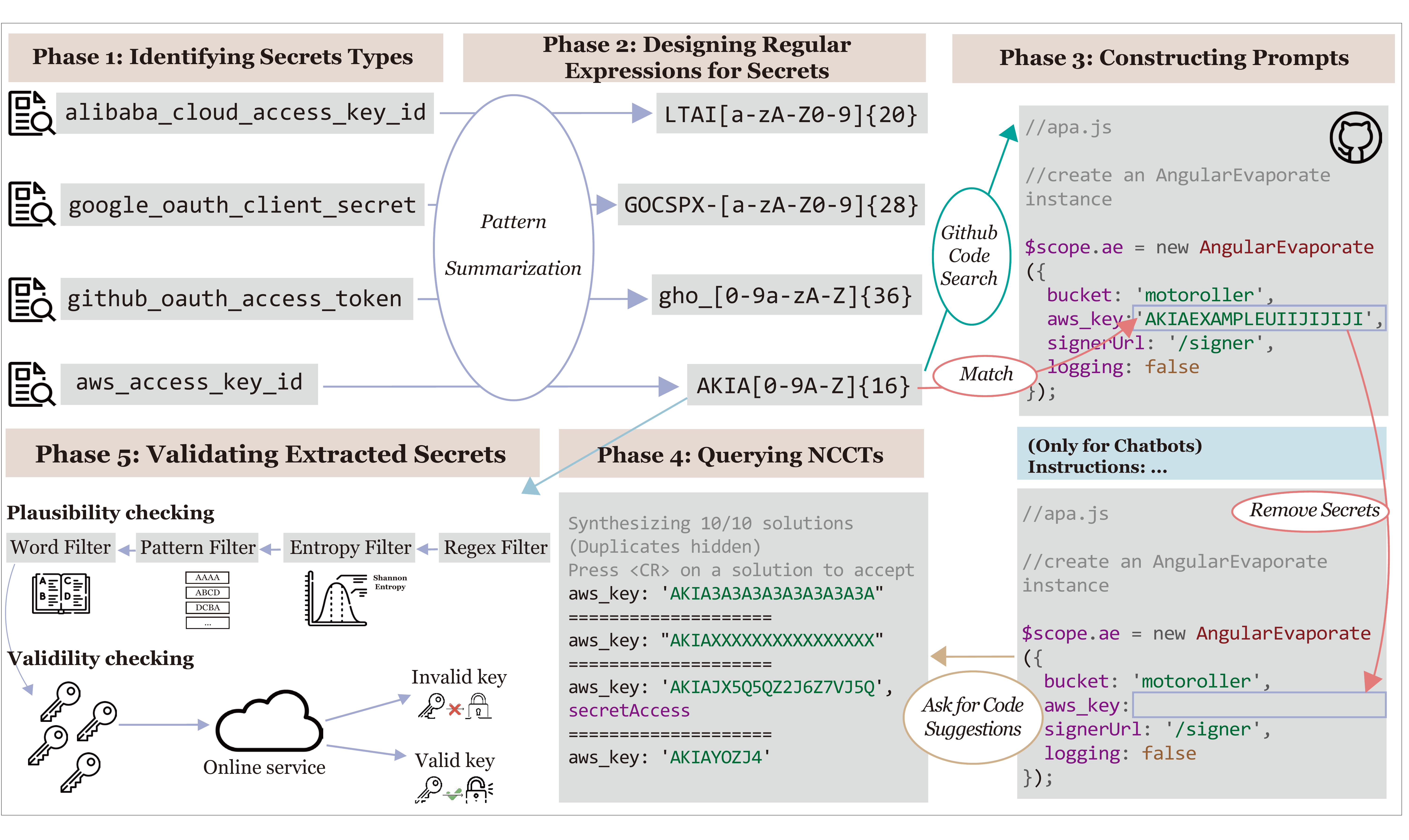}
\caption{The overall framework of attacking with HCR in five phases.}
\Description{The overall framework of attacking with HCR is in five phases: identifying secret types, designing regex for secrets, constructing prompts, querying NCCTs and validating extracted secrets. \textit{Note: as declared in Sec. \ref{ethics}, all secrets presented in this paper are rotated or masked, by authors and shown purely for demonstration purposes.}
}
\label{fig:hcr}
\end{figure*}
The design of HCR maximizes the chance of triggering memorization to reveal the risk of memorized hard-coded credentials. The key is to find effective prompts such that NCCTs will suggest plausible secrets. An effective prompt incorporates various aspects. These include details of the request to the online service, such as package imports, variable preparation, and the HTTP request main body. Therefore, we use code files with secrets from the most popular code hosting service --- GitHub. The intuitive idea is to use regular expressions (regex) to find the files with secrets. HCR collects files with secrets using regex from GitHub to construct prompts. Then, these prompts are used to query NCCTs. HCR further extracts secrets from the code suggestions of NCCTs and further validates them. Fig.~\ref{fig:hcr} illustrates the five phases of HCR with examples, and we elaborate the detailed design below.

\subsection{Phase 1: Identifying Secret Types}
\begin{table}[htbp]
\small
\caption{The collected secrets, their regex, the number of code files that matched our regex in GitHub, and their privacy risks. In the Risks column, the label ``D'', ``M'', and ``F'' are short for ``Data Breach'', ``Message Abuse'' and ``Financial Loss''. }
\centering
\label{regex}
\resizebox{\textwidth}{!}{%
\begin{tabular}{lllll}
\hline
\multicolumn{1}{c}{\textbf{Domain}} &
  \multicolumn{1}{c}{\textbf{Provider}} &
  \multicolumn{1}{c}{\textbf{Secret type}} &
  \multicolumn{1}{c}{\textbf{Regex}} &
  \multicolumn{1}{c}{\textbf{Risks}} \\ \hline
\begin{tabular}{l}
Social  \\
Media
\end{tabular} &
  Meta &
  facebook\_access\_token &
  EAACEdEose0cBA{[}0-9A-Za-z{]}+ &
  D,M \\ \hline
\multirow{3}{*}{\begin{tabular}{l}
Commu\\
-nication
\end{tabular}} &
  Slack &
  slack\_api\_token &
  \begin{tabular}[c]{@{}l@{}}xox{[}p|b|o|a{]}-{[}0-9{]}\{12\}-{[}0-9{]}\{12\}\\ -{[}0-9{]}\{12\}-{[}a-z0-9{]}\{32\}\end{tabular} &
  D,M \\ \cline{2-5} 
 &
  Slack &
  slack\_incoming\_webhook\_url &
  \begin{tabular}[c]{@{}l@{}}https:\textbackslash{}/\textbackslash{}/hooks.slack.com\textbackslash{}/\\ services\textbackslash{}/{[}A-Za-z0-9+\textbackslash{}/{]}\{44,46\}\end{tabular} &
  D,M \\ \cline{2-5} 
 &
  Sendinblue &
  sendinblue\_api\_key &
  xkeysib-{[}a-f0-9{]}\{64\}-{[}a-zA-Z0-9{]}\{16\} &
  D,M \\ \hline
\multirow{3}{*}{IaaS} &
  Alibaba Cloud &
  alibaba\_cloud\_access\_key\_id &
  LTAI{[}a-zA-Z0-9{]}\{20\} &
  D,F \\ \cline{2-5} 
 &
  \begin{tabular}[c]{@{}l@{}}Amazon Web \\ Services (AWS)\end{tabular} &
  aws\_access\_key\_id &
  AKIA{[}0-9A-Z{]}\{16\} &
  D,F \\ \cline{2-5} 
 &
  Tencent Cloud &
  tencent\_cloud\_secret\_id &
  AKID{[}0-9a-zA-Z{]}\{32\} &
  D,F \\ \hline
\multirow{3}{*}{SaaS} &
  Google &
  google\_api\_key &
  AIza{[}0-9A-Za-z\textbackslash{}-\_{]}\{35\} &
  D,F \\ \cline{2-5} 
 &
  Google &
  google\_oauth\_client\_id &
  \begin{tabular}[c]{@{}l@{}}{[}0-9{]}\{11,13\}-{[}a-z0-9{]}\{32\}\textbackslash{}.apps\textbackslash{}.\\ googleusercontent\textbackslash{}.com\end{tabular} &
  D,F \\ \cline{2-5} 
 &
  Google &
  google\_oauth\_client\_secret &
  GOCSPX-{[}a-zA-Z0-9{]}\{28\} &
  D,F \\ \hline
\multirow{5}{*}{Payment} &
  Midtrans &
  midtrans\_sandbox\_server\_key &
  SB-Mid-server-{[}0-9a-zA-Z\_-{]}\{24\} &
  D,F \\ \cline{2-5} 
 &
  Flutterwave &
  flutterwave\_live\_api\_secret\_key &
  FLWPUBK\_TEST-{[}0-9a-f{]}\{32\}-X &
  D,F \\ \cline{2-5} 
 &
  Flutterwave &
  flutterwave\_test\_api\_secret\_key &
  FLWSECK\_TEST-{[}0-9a-f{]}\{32\}-X &
  D,F \\ \cline{2-5} 
 &
  Stripe &
  stripe\_live\_secret\_key &
  sk\_live\_{[}0-9a-zA-Z{]}\{24\} &
  D,F \\ \cline{2-5} 
 &
  Stripe &
  stripe\_test\_secret\_key &
  sk\_test\_{[}0-9a-zA-Z{]}\{24\} &
  D,F \\ \hline
EC &
  eBay &
  ebay\_production\_client\_id &
  \begin{tabular}[c]{@{}l@{}}{[}a-zA-Z0-9\_\textbackslash{}-{]}\{8\}-{[}a-zA-Z0-9\_\textbackslash{}-{]}\{8\}-\\ PRD-{[}a-z0-9{]}\{9\}-{[}a-z0-9{]}\{8\}\end{tabular} &
  D,F \\ \hline
\multirow{2}{*}{DevOps} &
  GitHub &
  github\_personal\_access\_token &
  ghp\_{[}0-9a-zA-Z{]}\{36\} &
  D \\ \cline{2-5} 
 &
  GitHub &
  github\_oauth\_access\_token &
  gho\_{[}0-9a-zA-Z{]}\{36\} &
  D \\ \hline
\end{tabular}
}
\label{secret-type}
\end{table}

We find the list of secrets maintained by GitHub Secret Scanning Tool~\cite{gh_secret_scanning} useful. The list is a collection of many security-sensitive private keys, tokens, and APP IDs of many online services. We use 18 secret types in the list of GitHub Secret Scanning~\cite{gh_secret_scanning}, which are secrets of services of cloud computing, communications, payment, software development and operations (DevOps). Although 18 secret types are not exhaustive, they are representative and cover services widely used by software developers. We also list the potential risks if these credentials are leaked to an adversary in Tab.~\ref{secret-type}.
{
Note that to access to \textit{some} online services, paired credentials are required. However, the expose of even one of the paired credentials will significantly increase the risk of unauthorized access. For example, for Google OAuth Client ID, hard-coding itself is considered as safe, however, upon leakage of both client ID and client secret (which is also included in our study), this may result in unauthorised access to online services. }

\subsection{Phase 2: Designing Regular Expressions for Secrets}
\label{design-re}

In this phase, we need to design a regex that precisely describes the pattern for each type of secret, for instance, an AWS Access Key. 
{
This is feasible because online service credentials are usually designed to conform to certain regexes. Some online service providers even make the regexes for their credentials public, like Yandex~\cite{Yandex}.
Since most online service providers will not disclose the patterns of their secrets, we design the regexes for secrets by manual summarization. }Specifically, we generate ten secrets for each online service and summarize the observed pattern to design a concrete regex. 
{
Tab.~\ref{secret-type} shows the pattern. The pattern summarization is done by one of the paper's co-authors. After summarization, the patterns were verified by another co-author. 
The precision of the summarized pattern will be further validated in our experiments (see Sec.~\ref{validate_regex}).
}

Additionally, during implementation, we insert a regex \verb|[^a-zA-Z0-9_\-\/\\\+]| before and after each secret's regex pattern to \textit{exclude} prefix and suffix in the set of uppercase and lowercase letters (A-Z and a-z), digits (0-9), and special characters like hyphen (-), underscore (\_), forward slash (/), backslash (\textbackslash), and plus sign (+). This helps to eliminate potential false positive cases extracted by the secret's regex.

\subsection{Phase 3: Constructing Prompts}
\label{hcr_phase3}

The code files containing the secrets were collected by the GitHub Code Search~\cite{GitHubcodesearch}. This code search tool supports code searches with regexes, which is ideal for us to perform experiments. 
We collect files with secret types written in popular programming languages as defined in Tab.~\ref{secret-type}.
Besides, it is required that the computational cost to search with designed regex is reasonable, as it is observed that GitHub Code Search will crash with computationally expensive regexes. For instance, Mailchimp API Key exhibits the regex pattern ``\verb+[0-9a-f]{32}-us[0-9]{1,2}+'', but the only distinctive part is ``\verb+-us+'', which significantly increases the computational cost. GitHub Code Search denies high-cost regex searches because of excessive resource consumption.

For each candidate file with secrets, we spot the line where the first secret locates. We then delete the content in this line starting from the secret to EOL (End-Of-the-Line). We also detect and remove other existing secrets identified by regexes to eliminate the effect of context on memorization. Fig.~\ref{fig:hcr} shows an example: suppose originally the line is \verb+aws_key: 'AKIAIOSFQWER7EXAMPLE'+, after removing the secret, the line changes to \verb+aws_key: +. 
{

For general-purpose chatbots, we specify the instructions to perform the code completion task before the given code. See discussions in Sec. \ref{exp_setup} for details.
}

\subsection{Phase 4: Querying NCCTs}

After constructing the prompts in each file, we move the cursor to the first secret location and query the NCCTs to do code completion. 
After that, we save the NCCT outputs for further analysis.

\subsection{Phase 5: Validating Extracted Secrets}
\label{filter}

We are interested in whether the extracted secrets are \textit{plausible}. i.e., conform to the specific format. Beyond format checking, we also select a subset of credentials to perform \textit{validity} checking, by developing scripts to test whether the corresponding online service API grants our access to the online service. Due to ethical considerations, we \textit{will not} attempt to develop scripts for credentials that bring severe privacy concerns, e.g., API keys for live online payment. Details of ethical concerns are discussed in Sec. \ref{ethics}. In the following paragraphs, we elaborate on the plausibility checking.

We first extract the secrets in output with the regex described in Tab.~\ref{secret-type}.
Inspired by~\cite{meli2019bad}, we design three more filters for more fine-grained filtering. In all, we apply the four filters in sequence: the regex filter, the entropy filter, the pattern filter, and the word filter. We denote an extracted secret as \textit{plausible} if it passes all four filters.
The regular expression in the regex filter is the same regex as described in Tab.~\ref{secret-type}. Next, we discuss the implementations of the remaining three filters, which are the entropy filter, the pattern filter, and the word filter. 

\subsubsection{Entropy Filter}
A plausible secret shall exhibit a randomized pattern. Hence, we find an appropriate measurement of randomness in information theory is Shannon Entropy~\cite{shannon2001mathematical}. The Shannon entropy of a random variable is the average level of information,  or  ``surprise'' inherent to the variable's possible outcomes. Specifically, given a discrete random variable $X$, which takes a value $x$ in the alphabet $\mathcal{X}$ with probability $p$, the Shannon Entropy is 
\begin{equation}
\label{entropy_formula}
    H(X) := -\sum_{x \in \mathcal{X}} p(x) \log_2 p(x).
\end{equation}

In the context of credentials, higher entropy means a larger and more unpredictable range of possible values. Actually, in the industry practice, online service providers like GitHub~\cite{harvey_2021} are increasing the Shannon entropy of their credentials to defend brute-force attacks. Here, we denote a character in the string as the random variable $X$ and hence calculate the Shannon Entropy \textit{per character}. We use ``3-sigma rules'' to rule out candidate secrets that deviate from the average Shannon entropy per character with more than three standard deviations (std). {
Note that before calculating entropy, the fixed part of each secret is removed.

\textbf{Example calculation.} We illustrate how entropy filters work on an example candidate secret $s_1 =~$``AKIAXXXXXXXXXXXXBBBB".  First, we remove the fixed  ``AKIA'' part. In the remaining of the string, the occurrence frequency is 0.75, 0.25 for ``X'' and ``B'', respectively. Plugging this in to Eq.~\ref{entropy_formula}, we know that the Shannon entropy per character $H(s_1) = 0.8112$. Repeat the above entropy calculation for other candidate secrets $s_2, s_3, ..., s_n$, we get the average Shannon Entropy per character $\mu = \sum_i H(s_i) / n$, and std $\sigma=\sqrt{\frac{1}{N}\sum_i(s_i-\mu)^2}$. Suppose $\mu = 4, \sigma = 1$ in the population, we find that $H(s_1) = 0.8112 < \mu - 3\sigma = 1$,
so the secret $s_1$ falls outside the 3-sigma scope, hence we rule it out.}

\subsubsection{Pattern Filter}

The design motivation of pattern filters is that many false positive secrets exhibit trivial patterns.  Specifically, we design 6 patterns to rule out false positives. Denote literals \textit{A}, \textit{B}, \textit{C}, and \textit{D} as \textit{consecutive} characters, and denote \textit{W}, \textit{X}, \textit{Y}, and \textit{Z} as \textit{any} characters.

The filter detects six types of patterns in the secret string: 1. $AAAA$ 2.$ABCD$ 3.$DCBA$ 4.$XYXYXY$ 5. $WXYWXYWXY$ 6. $WXYZWXYZ$. For instance, the string $s_1$ in the above example matches pattern 1, so it is filtered out. To apply the word filter, we searched each candidate's secret for one of these patterns and failed to check if these pre-defined patterns were detected.

\subsubsection{Word Filter}
Apart from the non-trivial pattern, a plausible secret shall not contain words in natural language, for example, ``EXAMPLE'' and ``TESTING''. Such secrets are likely to be presented by developers as a showcase to use the secret. Therefore, we use a word dictionary containing the most common English words in top programming languages~\cite{anvaka2022}. We fail a candidate secret if it has a substring matching the word list.

We set the word filter length to 4 based on the F1-score results of our preliminary experiments on Tencent Cloud Secret ID for GitHub Copilot. That means if the length of the matched word in a candidate secret is shorter than 4, we still count the candidate secret as a plausible one. 

For some secret types, the fixed part of the secret contains English words, like the ``SB-Mid-server'' part in Midtrans Sandbox Server Key,  and the ``apps.googleusercontent.com'' part in Google OAuth Client ID. Therefore, we remove the fixed part of every secret before checking the word match in the candidate secrets.

An exceptional case is the eBay Production Client ID secret type, where the secret is designed to contain various English words names of people and places by nature,  for example ``AustinFr-ebayplug-PRD-xqlt71c7c-57qjc7xq''. Therefore we treat eBay Production Client ID as a special case and don't use the word filter to validate the secret type.

\section{Experiments}

This section first introduces the experimental settings and then evaluates HCR with the following Research Questions (RQs): 
\begin{enumerate}
    \item How effective is HCR on secret extraction?
    \item What are the characteristics of the memorized credentials?
    \item Can existing public code blocking methods defend against HCR?
\end{enumerate}
\subsection{Experimental Setting}
\label{exp_setup}
{

\textbf{Choice of NCCTs.}
We select a wider range of NCCTs, including commercial tools, open-source research models, and general-purpose chatbots with the capability of suggesting code snippets. 

For \textbf{commercial NCCTs}, We select two popular and representative ones:
}

(i) GitHub Copilot. Copilot is powered by OpenAI Codex with 12 billion parameters, and it is a descendant of GPT-3. Since its release in 2021, Copilot has become one of the most popular code completion tools, with VS Code Plugin downloads of over 7 million times. Copilot supports a code suggestion panel (by pressing ``Ctrl+Enter'') that provides up to 10 code suggestions by examining the code context, and these suggestions are ranked by the confidence score of the model~\footnote{The confidence score was available in the beta version of Copilot, as reported by~\cite{pearce2022asleep}. However, the scores have been removed since it was released to the public, so we are unable to collect the confidence score for further analysis.}. The version of the Copilot Plugin for the experiment is v1.92.177.  
    
(ii) Amazon CodeWhisperer. CodeWhisperer is a code LM developed by Amazon. It has more than 1.5 million times of downloads in VS Code Plugin. Different from GitHub Copilot, CodeWhisperer only supports one suggestion per query. In the experiment, the CodeWhisperer Plugin version is v1.77.0.  
{

For \textbf{open-source models}, to be consistent with settings in paper, where a NCCT may refer to context both above and below, we target models with the in-filling capability --- Stable Code and Code Llama. 

(iii) Stable Code. Stable Code~\cite{stable} is a decoder-only code LM for code completion that supports Fill in the Middle (FIM) capabilities and accommodates long context sizes. In our study, we employed the \textit{Stable Code-3B} version for experiments.

(iv) CodeLlama. CodeLlama~\cite{roziere2023code} is a code LM derived from Llama2~\cite{touvron2023llama} for code generation and infilling. After they have been pretrained on 500B code tokens, they are all fine-tuned to handle long contexts. Specifically, for this paper, we deployed the \textit{CodeLlama-13B} model for experiments.

For \textbf{general-purpose chatbots}, we select Gemini released by Google, and ChatGPT by OpenAI.

(v) Gemini. We conduct experiments on Gemini~\cite{gemini} by interfacing with its official API. Gemini is configured to its stable release version, specifically \textit{gemini-1.0-pro}. This setup is chosen to maintain consistency across different queries, aiming for reproducibility of results. 

(vi) ChatGPT. We call ChatGPT through the official API provided by OpenAI~\cite{gpt-3.5}, selecting the fixed version (\textit{gpt-3.5-turbo-0125}). This configuration ensures ChatGPT would generate the same output for the same query to ensure reproducibility.

\textbf{Experiment platforms.}
We run all the experiments of commercial NCCTs on a single PC with an Intel Xeon(R) W-2223 CPU, 32GB DDR4 RAM, using Windows 10. Due to the restricted usage of these commercial NCCTs, we use KeyMouseGo~\cite{anjianjingling}, a GUI Automation Tool to automate querying NCCTs. We tested all the NCCTs with their  Visual Studio Code plugins. For chatbots, we use the same PC to initiate the HTTPS request to the API of the chatbots.
We run the inference of open-source models on a server with Nvidia A100-PCIE Graphics card with 80GB GPU memory. 

Experiments of commercial NCCTs were conducted between 15 July and 1 August 2023, and experiments of open-source model and chatbots were conducted between 9 February and 1 March 2024.
}

\begin{table}[h]
\small
\centering
\caption{Distribution of programming languages in prompt construction.}
\begin{tabular}{|c|c|c|c|c|c|c|c|c|c|c|}
\hline
 & \textbf{JS} & \textbf{PHP} & \textbf{Python} & \textbf{TS} & \textbf{Java} & \textbf{Go} & \textbf{HTML} & \textbf{Ruby} & \textbf{C++} & \textbf{C} \\
\hline
\textbf{Count} & 208 & 202 & 194 & 78 & 65 & 52 & 48 & 46 & 4 & 3 \\
\hline
\textbf{Proportion} & 23.11\% & 22.44\% & 21.56\% & 8.67\% & 7.22\% & 5.78\% & 5.33\% & 5.11\% & 0.44\% & 0.33\% \\
\hline
\end{tabular}
\label{tab:pl_count}

\end{table}

\textbf{Prompt construction.}
For each of the 18 secret types, we first collect 80 files with secrets based on regex in HCR Phase 3. We then identify and drop the false positives extracted by the regex. After that, we randomly sample 50 files with secrets for each secret type. 
{
To craft the prompts, we first create a blank project for each extracted file with secrets. We then copied the extracted file that contains the secret to the project and removed the credential in the extracted file. Afterwards, we ask the NCCTs to complete the credential. Therefore, we have 18*50 = 900 pieces of prompts to query NCCTs in total. 

Tab.~\ref{tab:pl_count} shows the proportions of different programming languages in the final prompt set. Nine hundred pieces of files are from ten popular programming languages: JavaScript, PHP, Python, TypeScript, Java, Go, HTML, Ruby, C++, and C.  The line count in these code files ranges from a minimum of 4 to a maximum of 8158. We believe highly diverse prompts can comprehensively trigger secrets from NCCTs.

State-of-the-art general-purpose chatbots, like Gemini and ChatGPT, are usually trained to align with human values. This alignment serves to prevent the chatbot from producing harmful content, including hard-coded credentials. 
To break such safety alignment, our solution is to append the following English instructions to the chatbot prompts: 
\begin{tcolorbox}[width=\linewidth, boxrule=0pt, colback=gray!10, colframe=gray!20]
  Please replace [BLANK] in the following code with the appropriate string S to make the following code complete and runnable. You only output the string S. Do not explain your answer or make any changes to the rest of the code. $S$ must begin with ``X'' and must be Y characters long.

  <CODE>
\end{tcolorbox}

We mark the credential as a placeholder [BLANK], describe the format of the credential, and ask the chatbots to replace the placeholder without any explanation. In the above prompts, ``X'' is the prefix of the secret type, and ``Y'' is the length of the unfixed part of the secret.  $X$ and $Y$ are dependent on different secret types. <CODE> refers to the code file collected by HCR, with secrets replaced by a placeholder.

}

\subsection{RQ1: How Effective Is HCR on Secret Extraction?}

In this section, we first report the statistics of secret strings that pass the plausibility checking and then report the validity of the plausible strings.

\subsubsection{Plausibility of the Extracted Secrets}

\paragraph{Baseline methods.}
We select hand-crafted prompting as our baseline approach. Specifically, we craft \textit{five} distinct prompts for each credential category. This task was undertaken by one of the authors of this study. 

{
For each secret, we hand-craft the context of credentials, such as server URL and parameters required in the request body. Below, we show a Java example of the baseline method, where we ask NCCTs to complete line 1.
\lstset{
    commentstyle=\color{brown},
    keywordstyle=\color{magenta},
    numberstyle=\tiny\color{codegray},
    stringstyle=\color{codepurple},
    basicstyle=\ttfamily\footnotesize,
    breakatwhitespace=false,         
    breaklines=true,                 
    captionpos=b,                    
    keepspaces=true,                 
    numbers=left,                    
    numbersep=5pt,                  
    showspaces=false,                
    showstringspaces=false,
    showtabs=false,                  
    tabsize=2,
     numbersep=1pt,
    xleftmargin=10mm,
    xrightmargin=10mm
}
 
\begin{lstlisting}[caption=An example of hand-crafted prompt.]
String sendInBlue = 
String url = "https://api.sendinblue.com/v3/smtp/email";
String message = "Hello World";
JSONObject json = new JSONObject(message);
HashMap<String, String> header = new HashMap<>();
header.put("accept", "application/json");
header.put("content-type", "application/json");
header.put("api-key", sendInBlue);
\end{lstlisting}
}

\paragraph{Metrics.}
{

To report plausibility checking, we measure the effectiveness of HCR using \textit{Plausible Rate (PR)}. 
PR is defined as the ratio of \textit{the number of plausible secrets (PS\#) }to \textit{the number of total suggestions (TS\#)} for a NCCT. Note that sometimes NCCTs provide an empty suggestion, so the the actual number of suggestions may not reach the maximum suggestion number.
}

\paragraph{Experimental Results.}

\begin{table}[htbp]
\centering
\caption{The overall experimental result of extracted plausible secrets. ``PS\#'' means ``Plausible Secret Number'', ``TS\#'' means ``Total Secret Number'', and ``PR'' means ``Plausible Rate''.}

{

\begin{tabular}{c|c|c|c|c|c|c}
\hline
 & Copilot & CodeWhisperer & Code Llama & Stable Code & Gemini & ChatGPT \\
\hline
PS\# & 2702 | 103 & 129 | 24 & 100 | 4 & 52 | 2 & 445 | 179  & 96 | 42 \\
TS\# & 8127 | 781  & 736 | 89 & 194 | 7 & 70 | 2 & 551 | 212 & 353 | 149 \\
PR & 0.33 | 0.13 & 0.18 | 0.27 & 0.52 | 0.57 & 0.74 | 1.00 & 0.81 | 0.84 & 0.27 | 0.28\\
\hline
\end{tabular}
}
\label{tab:rq1_overall}
\end{table}

\begin{table}[htbp]
\caption{Comparison of Plausible Rate (PR) between HCR and hand-crafted prompting on commercial NCCTs. For each entry, the value before ``|'' denotes the results of HCR, while the value after ``|'' denotes the results of the baseline.}
\centering
\small
\begin{tabular}{l|ccc|c}
\hline
 & \multicolumn{3}{c|}{\textbf{Copilot}} & \textbf{CodeWhisperer} \\ \hline
& \textbf{Top-1} & \textbf{Top-3} & \textbf{Top-10} & \textbf{Top-1} \\ \hline
alibaba\_cloud\_access\_key\_id & 0.86 | 0.80 & 0.84 | 0.67 & 0.76 | 0.42 & 0.22 | 0.60 \\ \hline
aws\_access\_key\_id & 0.02 | 0.20 & 0.01 | 0.07 & 0.00 | 0.02 & 0.00  | 0.00\\ \hline
ebay\_production\_client\_id & 0.62 | 0.00 & 0.51 | 0.00 & 0.30 | 0.00 & 0.06  | 0.00\\ \hline
facebook\_access\_token & 0.02 | 0.00 & 0.01 | 0.00 & 0.00 | 0.00 & 0.00  | 0.00\\ \hline
flutterwave\_live\_api\_secret\_key & 0.74 | 0.40 & 0.71 | 0.33 & 0.53 | 0.11 & 0.19  | 0.00\\ \hline
flutterwave\_test\_api\_secret\_key & 0.60 | 0.00 & 0.53 | 0.00 & 0.32 | 0.00 & 0.10 | 0.20 \\ \hline
github\_oauth\_access\_token & 0.18 | 0.00 & 0.13 | 0.00 & 0.08 | 0.00 & 0.00  | 0.00\\ \hline
github\_personal\_access\_token & 0.45 | 0.00 & 0.38 | 0.00 & 0.27 | 0.00 & 0.34  | 0.00\\ \hline
google\_api\_key & 0.76 | 0.60 & 0.71 | 0.47 & 0.63 | 0.39 & 0.74 | 0.20 \\ \hline
google\_oauth\_client\_id & 0.54 | 0.20 & 0.50 | 0.14 & 0.37 | 0.06 & 0.52  | 0.00\\ \hline
google\_oauth\_client\_secret & 0.06 | 0.00 & 0.02 | 0.00 & 0.01 | 0.00 & 0.00  | 0.00\\ \hline
midtrans\_sandbox\_server\_key & 0.90 | 0.80 & 0.88 | 0.73 & 0.82 | 0.59 & 0.02  | 0.00\\ \hline
sendinblue\_api\_key & 0.20 | 0.20 & 0.14 | 0.15 & 0.06 | 0.05 & 0.05  | 0.00\\ \hline
slack\_api\_token & 0.08 | 0.00 & 0.06 | 0.00 & 0.03 | 0.00 & 0.00  | 0.00\\ \hline
slack\_incoming\_webhook\_url & 0.60 | 0.60 & 0.55 | 0.53 & 0.44 | 0.32 & 0.28 | 0.20 \\ \hline
stripe\_live\_secret\_key & 0.12 | 0.00 & 0.10 | 0.00 & 0.08 | 0.00 & 0.08  | 0.00\\ \hline
stripe\_test\_secret\_key & 0.84 | 0.80 & 0.79 | 0.60 & 0.67 | 0.23 & 0.65 | 0.40 \\ \hline
tencent\_cloud\_secret\_id & 0.62 | 0.00 & 0.60 | 0.00 & 0.50 | 0.00 & 0.00 | 0.00 \\ \hline \hline
avg. & 0.46 | 0.27 & 0.42 | 0.21 & 0.33 | 0.13 & 0.26 | 0.09\\ \hline
\end{tabular}
\label{commercial_result}
\end{table}

{
We report the overall experimental result of extracted plausible secrets in Tab.~\ref{tab:rq1_overall}. All six NCCTs are capable of generating a large amount of plausible secrets. Using HCR, Gemini has achieved the highest PR (0.81).

In Tab.~\ref{commercial_result}, we report the detailed secret extraction results of GitHub Copilot and Amazon CodeWhisperer.
We consider the top-1, top-3, and top-10 suggestions of GitHub Copilot. At the same time, CodeWhisperer only provides one suggestion per query, so we only report the secret extraction results based on the top-1 suggestions of Amazon CodeWhisperer.
}

There is an interesting finding when we use both HCR and handcraft prompting: examining the plausible rate of all suggestions of Copilot, we find that the plausible rate among top-$k$ suggestions increases when $k$ decreases. It suggests that Copilot has a higher confidence score when outputting a plausible secret.

\subsubsection{Validity of the Extracted Credentials}   
We have demonstrated that NCCTs generate a substantial amount of \textit{plausible} secrets that conform to their predefined format. 
We are further curious whether the extracted credentials are genuine \textit{valid} credentials that can pass the authentication of real online service APIs. Due to ethic considerations declared in Sec. \ref{ethics}, we \textit{will not} attempt to directly verify credentials that bring severe privacy concerns, e.g., API keys for live online payment. Instead, we select a subset of the credentials and develop validation scripts to probe whether the extracted credentials are valid. The validation scripts send requests to the corresponding online service APIs, and we check the status code in the response from APIs. 

The specific three types of secrets that we examine are Flutterwave Test API Secret Key, Midtrans Sandbox Server Key, and Stripe Test Secret Key. These keys are issued by online payment platforms for developers to test payment services in a sandbox. Therefore, validity checking on these credentials will not bring any practical privacy concerns. 

As for validation results, we managed to identify two valid Stripe Test Secret Keys using HCR, namely ``sk\_test\_******qLyjWDarjtT1******'' and ``sk\_test\_******JOvBiI2HlWgH******''. Key 1 was generated by five NCCTs, and key 2 was generated by four NCCTs. We further investigated the potential sources of these two keys, and evidences~\cite{stripe_doc_1,stripe_doc_2} show that these keys are two example secrets from official Stripe API documentation. Moreover, they are weakly memorized secrets which we will introduce in the next subsections. 

{

\begin{tcolorbox}[width=\linewidth, boxrule=0pt, colback=gray!20, colframe=gray!20]
\textbf{Answer to RQ1:}
HCR outperforms the baseline in prompting NCCTs to generate  plausible secrets and valid secrets. 
Moreover, GitHub Copilot has a higher confidence score when outputting a plausible secret.
Using HCR, we managed to identify two valid secrets that can be used to access the corresponding online service APIs. 
\end{tcolorbox}
}

\subsection{RQ2: What Are the Characteristics of the Memorized Credentials?}
\label{eva}
{

\subsubsection{Experimental Design}
We are further exploring whether these plausible credentials are \textit{memorized} secrets from NCCTs, i.e., members of the training dataset. Formally speaking, using the notations in Sec~\ref{bg-memo}, we suppose the training dataset is $D$. For a NCCT, we construct the prompt $[p_1||p_2]$, the concatenation of $p_1,p_2$, and get the NCCT output $s$. We are interested in whether $s\in D$.

To achieve this, we resort to GitHub Code Search to determine memorization\footnote{Note that the GitHub Code Search limits the identification memorized secrets. Hence, the empirical result in this study is \textit{a lower bound} of the real memorization rate.} --- if there is a search hit returned by the tool, due to the high Shannon entropy nature of the secret, the secret shall appear in the training data, and a NCCT memorizes it. We note that there might be chances that a NCCT generates a plausible secret without seeing it during training. Then it is called ``hallucinations''. But it is with nearly \textit{zero} probability that a hallucinated secret also happens to appear on GitHub.

Based on the properties of the prompt (the context) $p_1,p_2$, we further divide all the memorized secrets into two mutually exclusive categories: \textit{strongly}  and \textit{weakly} memorized secrets.  If $s\in D$ and $[p_1||s||p_2]\in D$, then $s$ is defined to be a \textbf{strongly memorized secret}. However, if $s\in D$ and $[p_1||s||p_2]\notin D$, then $s$ is defined to be a \textbf{weakly memorized secret}. In simple terms, for strong memorization, NCCTs memorize \textit{both} the secret and its context. In contrast, in weak memorization case, NCCTs \textit{only} memorize the secret. In practice, to determine whether $[p_1||s||p_2]\in D$, we just need to compare whether the removed secret in HCR Phase 3 matches the  secret suggested by a NCCT.

The potential appearance of weakly memorized secrets indicates that a NCCT might not solely reproduce the precise fragment of training data, but it could also inadvertently leak \textit{additional} secret strings. The existence of a weak memorization phenomenon poses a significant cyber-security risk since any secret string from training data might be inadvertently leaked.

We are interested in how many strongly / weakly memorized secrets are generated by each NCCT, and the corresponding proportion in all the plausible secrets. Therefore, we report the following metrics. 

\textit{Strongly Memorized Secret Number (SMS\#) and Strong Memorization Rate (SMR).} SMR is the ratio of the number of strongly memorized Secret (SMS\#) to \textit{the number of Plausible Secrets (PS\#)}.
\begin{equation}
\label{}
    SMR = \frac{SMS\#}{PS\#}.
\end{equation}

\textit{Weakly Memorized Secret Number (WMS\#) and Weak Memorization Rate (WMR).} Using GitHub Code Search, we obtain the  \textit{Memorized Secret Number (MS\#)}. Excluding the strongly memorized ones, we get WMS\#:

\begin{equation}
\label{}
    WMS\# = MS\# - SMS\#.
\end{equation}
The calculation of WMR is similar to SMR:
\begin{equation}
\label{}
    WMR = \frac{WMS\#}{PS\#}.
\end{equation}
Note the notation of strong and weak memorization are not applicable to the hand-crafted baseline methods. Therefore, 
\textit{Memorization Rate (MR)} is employed to show comparisons between HCR and the baseline.

\begin{equation}
\label{}
     MR = WMR + SMR
\end{equation}

\subsubsection{Experimental Results}

\begin{table}[htbp]
\small
\centering
\caption{Experimental Results of RQ2. ``PS \#'' means ``Plausible Secret Number'', ``MS \#'' means ``Memorized Secret Number'', ``MR'' means ``Memorized Rate'', ``SMS \#'' means ``Strongly Memorized Secret Number'', ``SMR'' means ``Strong Memorization Rate, ``WMS \#'' means ``Weak Memorization Secret Number'', and ``WMR'' means ``Weak Memorization Rate''. Entries with ``|'' shows the comparison between HCR (before ``|'') and hand-crafted prompting (after ``|'').}
{
\begin{tabular}{c|c|c|c|c|c|c}
\hline
 & Copilot & CodeWhisperer & Code Llama & Stable Code & Gemini & ChatGPT \\
\hline

PS\# & 2702 | 103 & 129 | 24 & 100 | 4 & 52 | 2 & 444 | 179 & 94 | 42\\ \hline
MS\#& 200 | 8  & 18 | 0  & 24 | 1 & 11 | 0 & 8 | 0 & 4 | 0\\
MR & 0.07 | 0.07 & 0.14 | 0.00 & 0.24 | 0.25 & 0.21 | 0.00 & 0.02 | 0.00 & 0.04 | 0.00 \\\hline
SMS\# &103 &11 & 11 & 2 & 3& 3\\
SMR &0.04  & 0.09 & 0.11 & 0.04 & 0.01 & 0.03 \\\hline
WMS\# &97 &7 &13 &9 & 5 & 1\\
WMR &0.04 & 0.05  & 0.13 & 0.17 & 0.01 & 0.01 \\
\hline
\end{tabular}
}
\label{tab:rq2}
\end{table}

Tab.~\ref{tab:rq2} reports the calculated metrics for each NCCT by applying both HCR and hand-crafted prompting.  Using hand-crafted prompting, it is observed that 4 out of 6 NCCTs have $MR=0$, i.e, do not suggest any memorized secrets. In contrast, using HCR, all NCCTs have non-zero SMR and WMR. This validates the superiority of HCR on effectively prompting NCCTs to generate memorized secrets.

Then, a nature question would be how to characterize the context that differentiates strong and weak memorization. We first consider two numerical variables to characterize the context $[p_1||p_2]$: \textit{the total line number of the code (``line\_num'')} and \textit{the total token number after tokenizing the code (``token\_num'').} Further considering $p_1$ and $p_2$ separately, we get four more variables, i.e., \textit{``line\_num\_above''
``token\_num\_above'', ``line\_num\_below'',} and \textit{``token\_num\_below''}. We encode the context into three types, hence getting a categorical variable with three possible values: 3 for context that leads to strong memorization, 2 for context that leads to weak memorization, and 1 for context that does not lead to memorization.

We use Mann–Whitney U test~\cite{MWU} method to test the statistical difference of  numeric values  within three groups. Compared to Student T test,  Mann–Whitney U test is free of the normal distribution assumptions, so it fits our data assumptions. 

We run Mann–Whitney U test with two representative NCCTs, GitHub Copilot and Amazon CodeWhisperer. During experiments, we use the tokenizer of Code Llama to tokenize all the code. The experimental result is shown in Fig. \ref{MWU}.

\begin{figure}[htbp]
\centering
\includegraphics[width=0.9\linewidth]{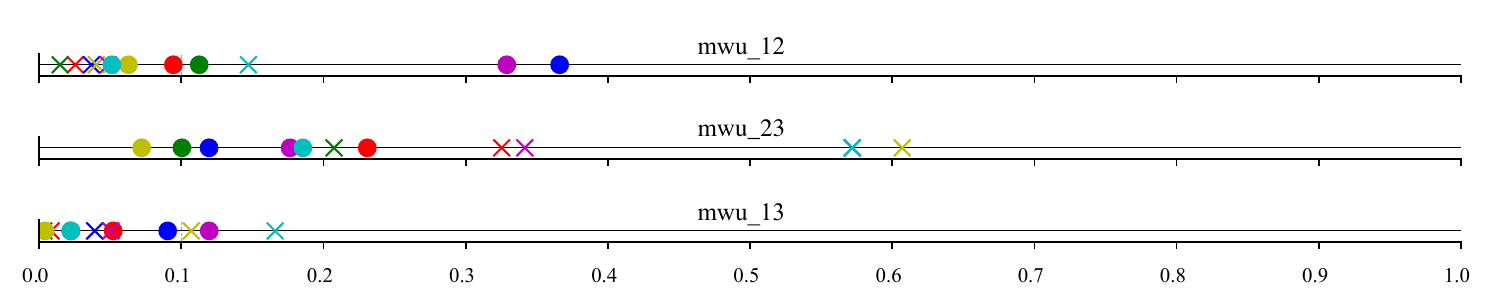}
\definecolor{r}{RGB}{251,22,31}
\definecolor{g}{RGB}{37, 125, 32}
\definecolor{b}{RGB}{0,50,248}
\definecolor{y}{RGB}{192,187,52}
\definecolor{m}{RGB}{184,40,185}
\definecolor{c}{RGB}{41,189,189}
\caption{We use different colors for \textcolor{r}{``line\_num''}, \textcolor{g}{``token\_num''}, \textcolor{b}{``token\_num\_above''}, \textcolor{y}{``token\_num\_below''}, \textcolor{m}{``line\_num\_above''}, and \textcolor{c}{``line\_num\_after''}.
`o' and `x' markers refer to Copilot and CodeWhisperer correspondingly. Specifically, for hypothesis test `mwu\_$XY$', we test whether variable value in type $X$ is \textit{smaller} than that in type $Y$.}
\Description{We use different colors for \textcolor{r}{``line\_num''}, \textcolor{g}{``token\_num''}, \textcolor{b}{``token\_num\_above''}, \textcolor{y}{``token\_num\_below''}, \textcolor{m}{``line\_num\_above''}, and \textcolor{c}{``line\_num\_after''}.
`o' and `x' markers refer to Copilot and CodeWhisperer correspondingly. Specifically, for hypothesis test `mwu\_$XY$', we test whether variable value in type $X$ is \textit{smaller} than that in type $Y$.}
\label{MWU}
\end{figure}

With p-value 0.10, we are confident to reject the null hypothesis in test `mwu\_12' and `mwu\_13' for most of the variables. This suggests that the ``line\_num`' and ``token\_num'' of context leading to \textit{weak} memorization are significantly smaller than that in \textit{strong} memorization. Additionally, the ``line\_num`' and ``token\_num'' of context leading to \textit{non-memorized} secrets are significantly smaller than that in \textit{strong} memorization.

\begin{tcolorbox}[width=\linewidth, boxrule=0pt, colback=gray!20, colframe=gray!20]
\textbf{Answers to RQ2}: Comprehensive experimental results validate that HCR prompts NCCTs to generate memorized secrets more effectively than the baseline. Moreover, we reveal that all tested NCCTs are capable of reproducing the precise piece of training data. Besides, NCCTs inadvertently leak additional secret strings, hence bringing severe privacy risks.
\end{tcolorbox}

}

\subsection{RQ3: Can Existing Public Code Blocking Methods Defend against HCR?}

\subsubsection{Experimental Design}
Some NCCTs provide users with the choice of blocking suggestions that match public code. Specifically, GitHub Copilot includes a filter that detects code suggestions matching public code on GitHub~\cite{copilot_coderef}. Users can choose to enable or disable the filter. With the filter enabled, GitHub Copilot checks code suggestions with their surrounding code of about 150 characters against public code on GitHub. The suggestion will not be shown to the user if there is a match or near match. Amazon CodeWhisperer also supports a similar functionality~\cite{aws_coderef}. Therefore, blocking suggestions that match public code may serve as an off-the-shelf defense against our HCR. 

{
However, we suspect that such internal matching algorithm is rather naive, since searching within the training dataset normally with $GB$ or $TB$-level size for $each$ user query is clearly not practical, due to large computational costs and significant latency. Therefore, such a mechanism could not detect strong memorization effectively. }
Since this feature does not take effect unless attackers opt-in, we also disable this feature in our previous experiments from the standpoint of attackers. Here, we examine whether existing public code blocking features can defend against our HCR when enforced compulsorily.

\begin{table}[htbp]
\caption{Experimental result comparisons between the public code blocking feature enabled (before ``|'') and disabled (after ``|'').}
\resizebox{0.7\textwidth}{!}{%
\begin{tabular}{ccccc}
\hline
 & \textbf{Plausible} \# & \textbf{Plausible Rate} & \textbf{SMR} & \textbf{WMR} \\ \hline
CodeWhisperer & 130 | 129 & 0.17 | 0.18 & 0.09 | 0.09 & 0.05 | 0.05 \\ \hline
Copilot & 1396 | 2702 & 0.29 | 0.33 & 0.02 | 0.04 & 0.03 | 0.04 \\ \hline
\end{tabular}
}
\label{public-code-match}
\end{table}

\subsubsection{Experimental Results}
Tab.~\ref{public-code-match} compares the experimental results when enabling and disabling the public code blocking features. For CodeWhisperer, we observe that the results remain nearly the same after the public code blocking feature is enabled. For GitHub Copilot, compared to the setting where the public code blocking feature is disabled,  48.3\% less plausible secrets are triggered after the public code blocking feature is enabled. SMR and WMR also drop by a small margin. 

\begin{tcolorbox}[width=\linewidth, boxrule=0pt, colback=gray!20, colframe=gray!20]
\textbf{Answers to RQ3}: As revealed by our experiments, even with the public code blocking feature enabled, our HCR can still obtain a considerable amount of plausible secrets. Besides, the number of strongly and weakly memorized secrets only decreases by a small margin. Therefore, it shows that the existing public code blocking method cannot effectively defend against our HCR. This confirms the distinct characteristics of secret leakage from NCCTs and calls for attention to this new privacy concern.
\end{tcolorbox}

\section{Method Validation}
In this section, we validate two key rule-based components in our proposed attack method HCR --- regexes and post-processing filters.

{
\subsection{Validation of Regexes}
}
For HCR, a loose regex will reduce the quality of collected code files in Phase 3 and increase the false positive number of extracted secrets in Phase 4. Therefore, it is pivotal to evaluate the precision of our regex. By manual inspection, we identify two types of false positive strings matched by our regex: 1) function / variable names; 2) the substring of a long string of random characters.

To assess whether the designed regex gives a tight bound to describe the pattern, for each secret type,  we draw a random sample of $n=100$ code files (or the number of hits returned by the GitHub Code Search, whichever is smaller) from the population of all code files containing that secret type. We then identify the false positives and report the precision. Unfortunately, the false negative number is unavailable due to the limitation of the GitHub Code Search. Hence, we are unable to report other metrics like recall.

For the validation result, the regex precision of most credentials is 1.0, and the average precision of all the regexes is 0.992. The only two summarized regexes not with 1.0 precision are the regex of eBay Production Client ID (0.87) and Stripe Test Secret Key (0.98). It shows that the summarized regex pattern is sufficient to work well in Phases 3 and 4.

\subsection{Validation of HCR Filters}

\label{validate_regex}
\begin{figure}[htbp]
\centering
\includegraphics[width=0.9\linewidth]{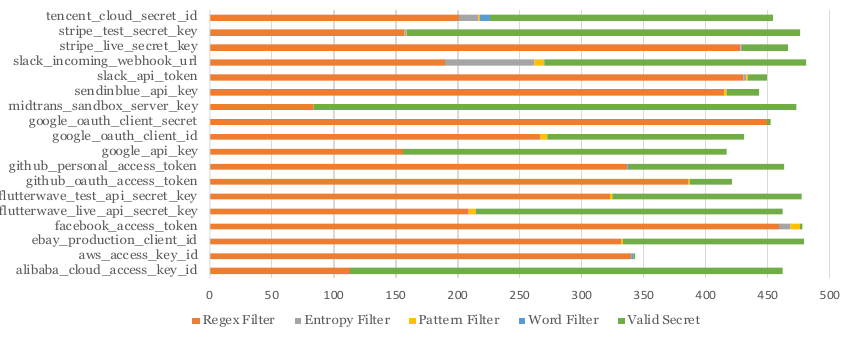}
\caption{The number of candidate secrets dropped by each filter applied in the order of the regex filter, the entropy filter, the pattern filter, and the word filter.}
\Description{The number of candidate secrets dropped by each filter applied in the order of the regex filter, the entropy filter, the pattern filter, and the word filter.}
\label{secret_and_filter}
\end{figure}
We showcase the filtering of candidate secrets generated by GitHub Copilot. 
After obtaining raw suggestions, we apply four filters one by one. The sequence is the regex filter, the entropy filter, the pattern filter, and finally, the word filter. Fig.~\ref{secret_and_filter} shows the number of secrets that fail the checking of each filter step by step. We detail the key statistics and insights for each filter below. 

\textbf{Regex filter.} Regular expressions filter out most of the invalid secrets. Upon examining the failure cases, we empirically identify three categories of suggestions that fail to pass the regex filter: 1) an empty string; 2) variable names from the code context, e.g., \verb|SECRET_ID|, or tries to get secret from environment variables, e.g., \verb|os.environ.get('SLACK_TOKEN')|; 3) a string whose format does not \textit{strictly} conform to the format as defined by regex.

\textbf{Entropy filter.} The average Shannon Entropy of all the secrets that pass the regex filter is 4.46, with a standard deviation of 0.71. By 3-sigma rules, all strings with entropy lower than 2.34 are filtered out. 

\textbf{Pattern filter.} Pattern filter filters out many secrets in Slack Incoming WebHook URL. Most failure cases of this secret type exhibit repeated patterns, e.g., ``xxxx'' and ``0000''.

\textbf{Word filter.}  The last filter we apply is the word filter. Among all the filtered words, we find that the most common filtered word is ``EXAMPLE''.

We managed to filter out over 60\% of Copilot's suggestions by applying these four filters. A majority of the invalid secrets were filtered out by the regex filter, the first filter that we applied. The filtering results differ significantly among various secret types, showcasing the effectiveness of the filters in identifying and filtering out secrets based on their distinct characteristics.

\section{Mitigation}

This section discusses potential mitigation strategies against the identified risk of NCCTs. From the perspective of programmers, we can reduce the occurrence of hard-coded credentials in code. Consequently, it reduces the possibility that the hard-coded credentials occur in the training data of NCCTs. From the perspective of NCCT developers, we can adopt different approaches at different stages of NCCT development. Before pre-training, we can exclude these hard-coded credentials from training by cleaning the training data; during NCCT training or fine-tuning, there are algorithmic defenses to guarantee privacy preservation; finally, during inference, we could post-process the NCCTs' output to filter out secrets.

\subsection{Eliminating Hard-Coded Credentials in Code} 
From the perspective of programmers, one possible solution is to employ centralized credential management tools.  Such tools are designed to eliminate hard-coded secrets by storing the secrets in a secured sandbox and accessing the secrets at runtime. 
Some popular centralized credential management tools include Microsoft Azure Key Vault~\cite{key_vault} and HashiCorp Vault~\cite{vault}.
Another solution is code scanning,  which scans and disallows committing code with hard-coded credentials.  Notable code scanning tools in the industry include Git-secrets~\cite{git_secrets} of AWS Labs.

\subsection{Cleaning Training Data}
\label{data_clean}
From the perspective of NCCT developers, they can add a dedicated data cleaning step when pre-processing training data to remove hard-coded credentials. Our proposed filter is readily applicable for this kind of training data cleaning. Note that training data cleaning should be done \textit{before} NCCT training. Therefore, it cannot make an existing NCCT secret-free. Instead, we shall perform training data cleaning and retrain the model from scratch. The total cost to mitigate would be the cost of data cleaning plus the retraining cost.
\subsection{Privacy-preserving Training}
Even with secrets in the dataset, we may use privacy-preserving training techniques to protect the privacy of a model.

Differential Privacy (DP)~\cite{dp_2008} is a method that offers strong guarantees of privacy of a model. We can say that a model $M$ is $\varepsilon$-differentially private if, for every possible output $x$, the probability that this output is observed never differs by more than $\exp(\varepsilon)$ between two datasets  $D$ and $D'$ that differ in only a single record.
In practice, DP-SGD (differentially-private stochastic gradient descent)~\cite{dp_2016} is a popular choice to achieve differential privacy for a model. DP offers a distinct advantage compared to training data cleaning as it does not need to design and maintain ad-hoc rules to filter out secrets.

Machine Unlearning~\cite{bourtoule2021machine,golatkar2020eternal} is a new idea to make ML models forget about particular data. The models receive a forget set $F$, a set of data points that need to be unlearned, and unlearning algorithms are required to eliminate the effect of the data points efficiently.
Although machine unlearning has not been applied to language models with billions of parameters, it is a promising direction to prevent NCCTs from leaking secrets.

\subsection{Post Processing}
Mitigation is possible at the NCCTs' inference stage without modifying the model itself. Specifically, we can detect and filter out secrets in NCCTs' output. This can be done by the filters of our HCR, as mentioned in Sec.~\ref{data_clean}. 
Unlike mitigation strategies discussed above, this method requires applying filters to NCCTs' outputs
\textit{per user query}. Although it saves the one-off cost of modifying the model in the short run, the computational cost can still be high considering accumulated user queries in the long run. Therefore, providers of NCCTs should consider this trade-off when choosing different mitigation strategies.

\section{Discussion}

\subsection{Threats to Validity}

\textbf{Secret type coverage.}
This work focuses on 18  secret types to demonstrate the effectiveness of our HCR and uncover the secret leakage issue of NCCTs. Although we cannot cover all the secret types in the world, we think that our conclusion can generalize well. The reason is that we select the studied secret types based on the well-established list of GitHub Secret Scanning~\cite{gh_secret_scanning}. The selected secret types also cover a broad spectrum of popular application domains, such as cloud computing services, communications, payment, software development and operations (DevOps). Therefore, the selected secret types are diverse and representative. Besides, we can conveniently extend our HCR to other secret types based on the method described in this work since secrets of other types also usually adhere to a specific pattern, making them identifiable via regexes.

\noindent
\textbf{Limitations of code search tools.} 
In this study, we acknowledge the limitations of our code search tool in identifying strongly or weakly memorized secrets in the suggestions of NCCTs. GitHub Code Search~\cite{GitHubcodesearch} disallows specific regex searches due to excessive resource consumption. Moreover, the search tool only supports code search on the default branch of a repository, so the search engine will not return results from other branches or historical versions of the code. Another popular code search tool we found is Google BigQuery \cite{GoogleBigQuery}. However, Google BigQuery only takes the snapshot of \textit{licensed} repositories. Therefore, the data covered by  Google BigQuery is less than that of GitHub Code Search. Therefore, we choose to employ GitHub Code Search in our study.

\subsection{Disclosure}
Our experiments of direct verification of credentials are conducted on limited types of credentials. Moreover, during direct verification, evidences~\cite{stripe_doc_1,stripe_doc_2} have shown that the two identified Stripe Test Secret Keys are example credentials in the Stripe official documentation. Their leakage will not pose any substantial security concerns. Hence, coordinated disclosure of the vulnerability is not necessary.

{

\subsection{Broader Implications of the Findings}
Our work reveals the memorization issue of NCCTs, demonstrating their ability to memorize and leak hard-coded credentials from their training data. Our findings have several broader implications for different  stakeholders.

For developers, if they hard-code credentials in their code, which is then used to train NCCTs, they suffer from a high risk of credential leakage due to the memorization issue of NCCTs. Therefore, developers should check whether their  code has been used as training data of NCCTs and take action to minimize the negative consequences of potential credential leakage. Besides, they should not hard-code credentials in their code in the future.

For the organizations that develop NCCTs, the memorization issue stresses their responsibility to conduct thorough privacy audits prior to the release of these NCCTs. This measure is crucial to safeguard against inadvertent privacy leakage caused by their NCCTs.

For the end-users of NCCTs, they should immediately report suspected credential leakage to relevant stakeholders during the interaction with NCCTs, which can help to protect the privacy of others and improve NCCTs.
}
\section{Related Work}
\label{ncc}

\subsection{Neural Code Completion}

A code completion tool can suggest a list of plausible completions for the next chunk of code. Powered by language models~\cite{intelliCode, black2021gpt, feng2020codebert, wang2021codet5, ChatGPT} neural code completion breaks the limitations of traditional code completion, which heavily relies on static analysis like type inference and variable name resolving. 

Numerous efforts~\cite{ black2021gpt, fried2022incoder, xu2022codeparrot, codegeex,   ChatGPT, wang2023codet5p,code_ai_chain,cat_lm, skcoder, li2024enhancing} have been dedicated to advancing automatic code completion by training language models adapted on code tokens. Neural code completion models often benefit from large-scale pre-training on massive codebases, which enables them to learn general programming patterns and language-specific syntax~\cite{chakraborty2022natgen,contrabert}. Fine-tuning on smaller, task-specific datasets helps tailor these pre-trained models to specific domains or coding styles~\cite{wang2021codet5, feng2020codebert, efficientft}. Furthermore, by incorporating surrounding comments or direct instructions (e.g., ChatGPT~\cite{ChatGPT}) as prompts, code completion models~\cite{ChatGPT, codegeex, copilot_doc} can generate more precise code snippets, resulting in improved performance and accuracy~\cite{wang2022no}, thereby delivering a more tailored and practical coding experience for developers.

\subsection{Concerns of Neural Code Completion}
Several research works raise the concerns of neural completion tools. \citet{mastropaolo2023robustness} propose to test the robustness of GitHub Copilot. They found that with different but semantically equivalent natural language descriptions~\cite{zhang2023towards}, Copilot may suggest different recommendations. \citet{rabin2023study} study how neural code completion models are distracted by features of code. \citet{probing}
tested the robustness of the numerical and logical reasoning abilities of these models. 
A short paper discusses how using open-source code without sanitization brings privacy risks~\cite{al2023ab}. The work indicates that these neural code completion tools may leak private information, including credentials, API keys, directory structures, and in-code discussions by
developers. ~\citet{copilot_recitation} conducted a related work investigating GitHub Copilot's training data recitation. The experimental result shows that Copilot rarely recites its training data in an internal trial of Copilot in its infancy. Our work instead shows that attackers can deliberately trigger NCCTs like Copilot to recite hard-coded credentials in their training data.

{
In summary, different from previous studies, our work proposes an effective tool to evaluate the memorization issue of language models in the code domain, demonstrating their ability to memorize and leak hard-coded credentials from their training data. Besides, our method does not require full access to the training data of NCCTs, which enables us to uncover both strong and weak memorization of NCCTs. 
}

\section{Conclusion and Future Work}

{
This study demonstrates a new privacy risk of NCCTs by designing and implementing an effective tool, HCR, to extract hard-coded secret strings from 6 NCCTs. }In the proposed HCR, we constructed 900 prompts to query NCCTs to complete 18 types of secrets.
The experimental result reveals that with proper prompts, NCCTs may inadvertently leak secret strings.

The same attack strategy can generalize to any neural code completion tools, and developers of these tools shall consider auditing their NCCTs to determine the privacy level they offer and employ mitigation measures before releasing to the public. 

There are many open questions that we want to further explore in the future, including investigating password leakage from NCCTs, and  countermeasures to prevent secret leakage that are both effective and efficient. We hope this study raises the privacy concerns of the wide deployment of language models, when the privacy risk of language models is quite under-explored.

\begin{acks}
We thank all the anonymous reviewers for their invaluable comments and suggestions. 
This work was supported by the National Natural Science Foundation of China (Grant No. 62206318) and the Research Grants Council of the Hong Kong Special Administrative Region, China (No. CUHK 14206921 of the General Research Fund).
\end{acks}

\bibliographystyle{ACM-Reference-Format}
\bibliography{sample-base}

\end{document}